\documentclass[twocolumn,superscriptaddress,showpacs,amsmath,amssymb,pra,longbibliography]{revtex4-1}
\usepackage[pdftex]{graphicx} 
\usepackage{dcolumn}  
\usepackage{bm}       
\usepackage[usenames,dvipsnames]{color}
\definecolor{URLCOL}{rgb}{0,0.52,0.83} 
\definecolor{LINKCOL}{rgb}{0.05,0.5,0} 
\definecolor{orange}{rgb}{0.6,0.3,0} 
\definecolor{CITECOL}{rgb}{0.25,0,0.48} 
\usepackage{epstopdf}

\definecolor{darkgreen}{rgb}{0.0, 0.5, 0.0}
\graphicspath{{../f/}}

\usepackage[pdftex,bookmarks,breaklinks,bookmarksopen,bookmarksnumbered,colorlinks,linkcolor=LINKCOL,linktocpage,citecolor=CITECOL,urlcolor=URLCOL,pdfpagemode=UseOutline,pdftex]{hyperref}

\usepackage{booktabs}
\usepackage{comment}
\usepackage{natbib}

\definecolor{TITLECOL}{rgb}{0.1,0.2,0.7} 
\definecolor{SECOL}{rgb}{0.1,0.2,0.7} 
\definecolor{CONTENTSCOL}{rgb}{0.1,0.2,0.7} 
\definecolor{SSECOL}{rgb}{0.25,0,0.48} 
\definecolor{SSSECOL}{rgb}{0.2,0.08,0.53} 
\definecolor{FINCOL}{rgb}{0.01,0.3,0.07} 

\def\coloredtitle#1{\title{\textcolor{TITLECOL}{#1}}} 
\def\coloredauthor#1{\author{\textcolor{CITECOL}{#1}}} 

\definecolor{URLCOL}{rgb}{0,0.17,0.43} 
\definecolor{LINKCOL}{rgb}{0.05,0.4,0} 
\definecolor{CITECOL}{rgb}{0.35,0,0.48} 
\def\sss{\scriptscriptstyle\rm}

\def\bea{\begin{eqnarray}}
\def\eea{\end{eqnarray}}
\def\ben{\begin{equation}}
\def\een{\end{equation}}
\def\benu{\begin{enumerate}}
\def\enu{\end{enumerate}}

\def\bei{\begin{itemize}}
\def\eei{\end{itemize}}
\def\beit{\begin{itemize}}
\def\eit{\end{itemize}}
\def\benu{\begin{enumerate}}
\def\enu{\end{enumerate}}

\def\half{\frac{1}{2}}

\def\F{_{\sss F}}

\def\TF{^{\rm TF}}

\def\n{n}

\def\eps{E}

\def\sc{^{\rm sc}}

\DeclareMathOperator{\Ai}{Ai}

\begin{document}

\coloredtitle{%
Corrections to Thomas-Fermi densities at turning points and beyond
}

\coloredauthor{Raphael F. Ribeiro}
\affiliation{Department of Chemistry, University of California, Irvine, CA 92697}
\coloredauthor{Donghyung Lee} 
\affiliation{Samsung SDI Inc., SMRC, Samsung-ro 130, Yeongtong-gu, Suwon-si,
Gyeonggi-do, Republic of Korea, 443-803}
\coloredauthor{Attila Cangi}
\affiliation{Max Planck Institute of Microstructure Physics, Weinberg 2, 06120 Halle (Saale),Germany}
\coloredauthor{Peter Elliott}
\affiliation{Max Planck Institute of Microstructure Physics, Weinberg 2, 06120 Halle (Saale),Germany}
\coloredauthor{Kieron Burke}
\affiliation{Department of Chemistry, University of California, Irvine, CA 92697}

\date{\today}

\begin{abstract}

Uniform semiclassical approximations
for the number and kinetic-energy densities are derived for many non-interacting fermions in one-dimensional potentials with two turning points.
The resulting simple, closed-form expressions contain the leading corrections to Thomas-Fermi theory, involve
neither sums nor derivatives, are spatially uniform approximations, and are exceedingly accurate.

\end{abstract}


\maketitle
Semiclassical approximations are both ubiquitous in physics \cite{BB03, Ch91} and
notoriously difficult to improve upon.  Most of
us will recall the chapter on WKB in our quantum textbook\cite{GB05},
yielding
a simple and elegant result for the eigenvalues of a particle
in a one-dimensional potential.  The more sensitive will have recoiled at
the surgical need to stitch together various regions (allowed,
turning point, and forbidden) to find the semiclassical eigenfunction.
Summing the probability densities in the allowed region yields the dominant
contribution to the density, but what are the leading corrections?

A little later, we should have learned
Thomas-Fermi (TF) theory\cite{T27,F27}.
Thomas derived what we now call the TF equation
in 1926, without using Schr\"odinger's equation\cite{S26}.  
He calculated the energies of atoms, finding results accurate
to within about 10\%.  
TF theory has since been applied in almost all areas of physics\cite{Sa91}.
For the electronic structure of everyday matter, TF theory is
insufficiently accurate for most purposes, but gave rise to
modern density functional theory (DFT)\cite{HK64}.
The heart of TF
theory is a local approximation, and the success of
semilocal approximations in modern
DFT calculations of electronic structure can be
traced to the exactness of TF in the semiclassical limit\cite{LS73,L81}.
So, what are the leading corrections?

Despite decades of development in quantum theory,
the above questions, which are
intimately related, remain unanswered.  
Both the WKB and the TF approximations can be derived from any formulation of
non-relativistic quantum mechanics, but none yields an obvious
procedure for finding the leading corrections.
Mathematical difficulties arise because $\hbar$ multiplies the highest derivative in the
Schrodinger equation. Physically, the problem is at
the dark heart of the relation between quantum and classical mechanics.

\begin{figure}[htb]
\includegraphics[scale=0.5]{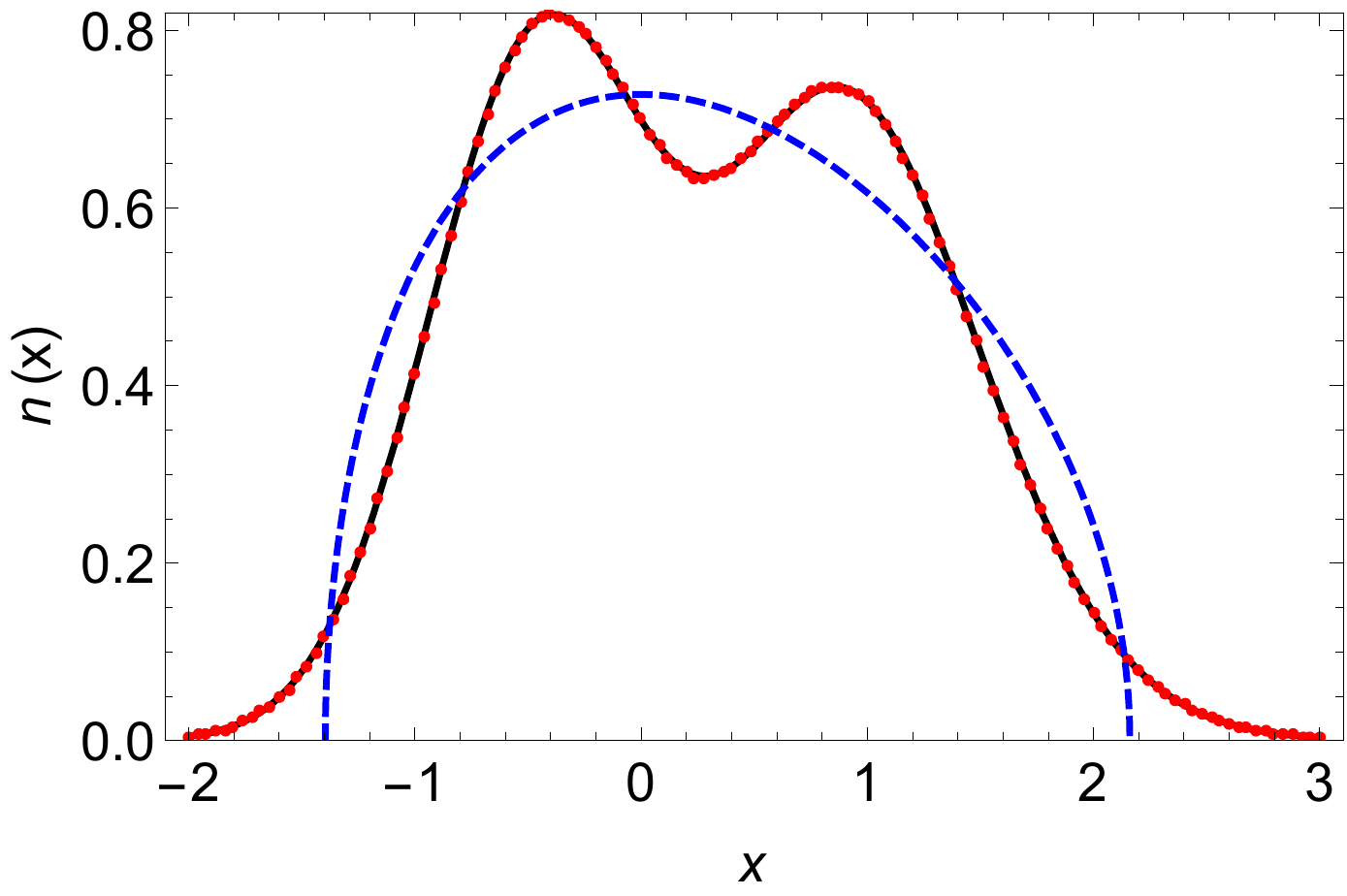}
\caption{Thomas-Fermi (dashed) and semiclassical (dotted) approximations
to the density (solid) 
of 2 particles in a Morse potential, $v(x) =15 (e^{-x/2}-2\, e^{-x/4})$.}
\label{Morse20D}
\end{figure}

Here we derive a definitive solution to both these questions
in a limited context: Non-interacting fermions in one dimension.
Researchers from solid-state, nuclear, and chemical
physics have sought this result for over 50 years
\cite{Al61, SZ62, Pa63, Pa64, KSb65, Gr66,  BZ73, LY72, LL75, E88, ELCB08}.
The TF density for the lowest $N$ occupied orbitals is
\ben
\n\TF(x) = p\F(x)/(\hbar\pi),~~~~~~~p\F(x) \geq 0
\label{ntf}
\een
where $p\F(x)$ is the classical momentum at the Fermi energy, $E\F$,
chosen to ensure normalization, and vanishes elsewhere.
This becomes
\begin{widetext}
\begin{equation}
n\sc(x)= \frac{p\F(x)}{\hbar}\left[ \left(\sqrt{z} \Ai^2(-z) +
\frac{\Ai^{'2}(-z)}{\sqrt{z}} \right) +
\left(\frac{\hbar\omega_F \text{csc}[\alpha_F(x)]}{p\F^2(x)} 
- \frac{1}{2 z^{3/2}}\right) \Ai(-z)\, \Ai'(-z)\right]_{z=z\F(x)},
\label{nsc}
\end{equation}
\end{widetext}
where $p\F(x)$ is analytically continued into evanescent regions,
$\omega\F$ is the classical frequency at $E\F$,
and $z\F(x)$ and $\alpha\F(x)$ are related to the classical action from the
nearest turning point, and $\Ai$ and $\Ai'$ are the Airy function and its
derivative (details within).
Eq. (\ref{nsc}) contains the leading corrections to Eq. (\ref{ntf}) for 
{\em every} value of $x$, without butchery at the turning points.
The primary importance of this work is the {\em existence} of Eq. (\ref{nsc}) and
its derivation.
A secondary point is the sheer accuracy of  Eq. (\ref{nsc}):
For $N > 1$, its result is usually {\em indistinguishable} (to the eye)
from exact, as in Fig. \ref{Morse20D}. 
Generalization of Eq. (\ref{nsc})
could prove invaluable in any field using semiclassics or in orbital-free DFT\cite{CGB13}.

The crucial step in the derivation is
the use of the Poisson summation formula\cite{MF1, Cr79}. 
While long-known\cite{BM72, BT76, Cr79} for the
description of semiclassical phenomena, it has been little applied to bound states.
Although the bare result of its application appears quite complicated,
each of the resulting terms, which include contributions from every closed classical
orbit at the $E\F$, can be simplified and summed.
We assume only that the potential $v(x)$ is slowly-varying
with dynamics lying on a topological circle. 
Accuracy improves as the number of particles grows except when
$E\F$ is near a critical point of $v(x)$.

To begin,
at energy $\eps$, the left ($x_-$) and right ($x_+$) classical
turning points satisfy $v(x_\pm)=\eps$.
The action, measured from the left turning point, is
\ben
S(x,\eps) = \int^x_{x_-(\eps)} dx\, p(x,\eps)
\een
where $p(x,\eps)=\sqrt{2 m [\eps-v(x)]}$ is the classical momentum.
The WKB quantization
condition \cite{K26, BM72,Ch91} is then
\ben
S\left[x_+(\eps_j),\eps_j \right] = \pi\hbar\left(j+\half\right),~~~~~j \in \mathbb{N}.
\label{EWKB}
\een
The accuracy of WKB quantized energies
generally improve as either $j$ or $m$ grows, $\hbar$ shrinks, or the
potential is stretched such that its rate of change becomes smaller \cite{Ch91, BO78}.
But the WKB wavefunction is
singular in the turning point region
\cite{J24,W26, B26, K26, Ch91}. 
Langer \cite{La37} obtained a semiclassical 
wavefunction
for the case where turning points are simple zeroes of the momentum:
\ben
\phi_j(x)=\sqrt{\frac{2m\omega_j }{p_j(x)}}\, 
z_j^{1/4}(x)\, \Ai\left[-z_j(x)\right],
\label{phiLang}
\een
where
$ \omega_j=\hbar^{-1} \partial E_\lambda /\partial \lambda|_{\lambda=j}$ 
is the frequency of the corresponding classical 
orbit, and
$z_j=  \left[3S_j(x)/2\hbar\right]^{2/3}$.
In a classically-forbidden region, $-p(x) =-i|p(x)| =
e^{3i\pi/2} |p(x)| $, ensuring continuity through the turning point.
The Langer solution can also be used for problems with two turning points \cite{M68}.
In this work we match Langer functions from each
turning point at the mid-phase point $x_m^j$ where $S_j(x_m^j)=\hbar (j+1/2) \pi/2$. This procedure ensures continuity everywhere.

Our task is to use Langer orbitals to find the asymptotic
behavior of the density of $N$ occupied orbitals,
\ben
n(x) = \sum_{j=0}^{N-1} |\phi_j(x)|^2.
\label{ddef}
\een
We use the
Poisson summation formula:
\ben
\sum_{j=0}^{N-1} f_j =
\sum_{k=-\infty}^{\infty} \int_{-1/2}^{N-1/2} d\lambda\, 
f(\lambda) e^{2\pi i k \lambda}, 
\label{pdef}
\een
where $f(\lambda)$ is essentially any continuous function with bounded first derivatives (except for a finite number of points) that matches the $f_j$ when $\lambda \in \mathbb{N}$~\cite{MF1, Be66, Cr79}.  
Write
\ben
n(x) = n_0(x)+n_1(x),
\label{ndef}
\een
where $n_0(x)$ is the contribution from $k=0$, and $n_1(x)$ is all the rest.
Then, for $m=1$,
 
\ben
n_0(x) = 2\int_{-1/2}^{N-1/2}\mathrm{d}\lambda  \frac{ \omega_\lambda \sqrt{z_\lambda (x)}}{p_\lambda (x)} \Ai^2[-z_\lambda(x)] .
 \label{n0def}
 \een
The lower bound of the integral corresponds to the stable fixed
point of the potential well, and the upper bound defines $E_F$
as that obtained by solving Eq. (\ref{EWKB}) \ for $j=N-1/2$, 
where $N$ is the number of particles in the system.  
Hereinafter, a subscript $F$ denotes evaluation at $E\F$, and
$x$ is treated as a parameter. For instance, to approximate the integral in Eq. \ref{n0def} we employ the transformation $\lambda \rightarrow p_\lambda(x)$.
Integrating by parts, using the Airy differential equation \cite{VS04},
changing variables, and neglecting higher-order terms from the lower-bound of the integral in Eq. \ref{n0def}, we find:
\ben
n_0(x) \sim \hbar^{-1} p\F(x)\,g_+[z\F(x)]+
\int_{z_{-1/2}(x)}^{z_F(x)} \mathrm{d}z\, {\sqrt{z}} \frac{\partial f}{\partial z} 
g_-(z),
\label{n0new}
\een
where
\ben
g_{\pm} (z)=z^{1/2}\, \Ai^2(-z) \pm z^{-1/2}\, {\Ai'}^2(-z)
\label{hdef}
\een
$f(z)=p(z)/{\sqrt{z}}$, and $\Ai'(z) = \mathrm{d}\Ai(z)/\mathrm{d}z$.

Eq. (\ref{n0new}) is useful for the extraction of the dominant terms in
an asymptotic expansion for $n_0(x)$.  As $N$ grows, the coefficients $\sqrt{z} \partial f/\partial z$ become ever more 
slowly-varying functions of the energy.
Integrating by parts, ignoring the remaining higher-order
contribution, and using
\ben
\frac{\partial f}{\partial z}\bigg |_{E_F,x}= \frac{\omega\F}{p\F(x) \alpha\F (x)} 
- \frac{p\F(x)}{2 \hbar z\F^{3/2}(x)},
\label{dfdzdef}
\een
where
$\alpha_F(x) =$ 
$\sqrt{z_F(x)} \hbar^{-1}~ \partial z_\lambda(x) /\partial \lambda |_{\lambda=N-1/2}$
(e.g, $ =  \omega_F \int_{x_-(E_F)}^x \mathrm{d}x'/p(x')$ for $x_{-}(E_F)< x < x_m$). We find
\ben
n_0(x) \sim \hbar^{-1} p_F(x) g_+[z\F(x)] +\frac{\partial f}{\partial z} \bigg|_{E_F,x}\ A_0[z\F(x)],
\label{n0semi}
\een
where
$ A_0(z)= \Ai(-z) \Ai'(-z)$.

To evaluate the $k\neq 0$ components of Eq. \ref{pdef},
we use the integral representation of $\Ai^2(-z)$ \cite{VS04}
and change variable 
to $G_\lambda(x)= 2\pi\, k\, \lambda - z_\lambda(x)\ t$,
\ben
n_{1}(x) =
2\sum_{k=-\infty}^{\infty'}\small{\lim_{T \rightarrow \infty}} \int_{-T}^{T}
\mathrm{d}t~ \kappa(t)
\int_{G_{-1/2}}^{G_F} \frac{\mathrm{d}G_\lambda~\omega_\lambda \sqrt{z_\lambda}}{p_\lambda \frac{\partial G_\lambda}{\partial \lambda}} e^{iG_\lambda},
\label {n1int}
\een 
where the sum is over all $k \neq 0$, and $\kappa=i\exp(it^3/12)/(4{\sqrt{i\pi^3 t}})$. 
Integration by parts assuming negligible contributions from the lower bound 
yields, to leading order 
in $\hbar$ (or $1/N$):
\ben
n_{1}(x) \sim 2 \frac{\omega_F \sqrt{z_F}}{p_F} \sum_{k=-\infty}^{\infty'} (-1)^k  \lim_{T \rightarrow \infty}  \int_{-T}^{T}
\mathrm{d}t \frac{\kappa(t)e^{-i z_F t}}{2\pi k - y_F t},
\label{n1int2}
\een
where $y_F=\alpha\F/{\sqrt{z\F}}$. The factor $(2\pi k - y_F t)^{-1}$ may be expressed
as geometric series in $ t y_F/(2\pi k)$, with a radius of convergence 
$R_F = \left| 2 \pi k/y_F \right |$, which becomes arbitrarily large as $|k|$ becomes greater
and $|y_F|$ becomes smaller.  
This condition is generally fulfilled when $v(x)$ has an infinite number of
bound states, or if the semiclassical limit is
approached by stretching the coordinate\cite{ELCB08, CLEB10,L81}.
Assuming any errors introduced by this sequence of operations vanish in the semiclassical limit
the integrals required for the evaluation of $n_1(x)$ can be performed \cite{VS04}
and the results summed to give an asymptotic expansion for $n_1(x)$ in terms of
$A_0[z] = \Ai[-z]Ai'[-z]$, $A_1[z]= \Ai^2[-z]$ and $A_2[z] = \Ai^{'2}[-z]$:
 \ben
 n_1(x) \sim \frac{\omega_F}{p_F}\sum_{p=0}^{2} \sum_{j=0}^{\infty}\left(-z_F\right)^{-3j -p} \xi_{3j+p} (\alpha_F)A_p[z_F], 
 \een
where $\{\xi_j(\alpha_F)\}$ correspond to different power series in $\alpha_F(x)$, e.g.,
\ben
\xi_0(\alpha) = \sum_{k=1}^{\infty} \frac{(-1)^{k-1} 2 \left(2^{2k-1} -1\right) B_{2k}}{(2k)!} \alpha^{2k-1},
\label{csc}
\een
where $B_{2k}$ denotes the $2k$th Bernoulli number \cite{AS72}. Eq. \ref{csc} may also be expressed as $- 1/\alpha + \text{csc}~ \alpha $.
However, to extract the leading term of $n_1(x)$, only the term
with highest-power in $z_F(x)$ needs to be considered, yielding
\ben
n_1(x) \sim \frac{\omega_F}{p_F(x)}\left[\text{csc} \left[\alpha_F(x)\right]- \frac{1}{\alpha_F(x)}\right]A_0[z_F(x)].
\label{n1semi}
\een
The sum of Eqs. \ref{n0semi} and \ref{n1semi} yields Eq. (\ref{nsc}).
The relative orders of each term in $\hbar$ only become explicit after
accounting for the $z_F(x)$ dependence, which changes in different regions (see below).
For instance, while the rightmost term in 
Eq. \ref{nsc} has a multiplying factor of $\hbar^{-1}$, it is canceled by the $\hbar^{-1}$ in $z_F^{-3/2}(x)$.
Equation \ref{nsc} also illustrates the vital balance between the asymptotic expansions constructed for $n_0(x)$ and $n_1(x)$.
The former (see Eq. \ref{n0semi}) contains the pole 
$ \alpha_F^{-1}$ of the Laurent series for $\text{csc} \left(\alpha_F \right)$ about $\alpha_F = 0$ (turning point), whereas Eq. \ref{csc} contains all remaining terms of the series.

Further, if we choose 
\ben
t(x) = \sum_{j=0}^{N-1} p_j^2(x) |\phi_j(x)|^2 /2,
\label{tdef}
\een
similar steps produce
\begin{align} &t\sc(x) =  \frac{p\F^2(x)}{6}\, \n\sc(x) + 
\frac{p_F(x)\omega_F}{3\text{sin}\alpha_F(x)}A_0[z_F(x)].
\label{tsc}
\end{align}
Eqs. (\ref{nsc}) and (\ref{tsc})
define closed form global uniform semiclassical approximations
to $n(x)$ and $t(x)$ which
are asymptotically exact
as $\hbar \rightarrow 0$ or $N\to \infty$.

These approximations simplify in different regions.

\noindent{\em Classically-allowed:}
For $z_F(x) >> 1$, the asymptotic form of the Airy function applies, leading to
\ben
\n\sc(x) \to
\frac{p_F(x)}{\hbar\pi} - \frac{\omega_F \cos\left[2S_F(x)/\hbar\right]}{2\pi p_F(x) \sin\alpha_F(x)},
\label{clallow}
\een
(simplifying Eq. (3.36) of Ref. \cite{KSb65}; see also \cite{LL75}).
The dominant smooth term arises from the direct short-time
classical orbit\cite{BM72, LY72}.
The oscillatory contributions arise from single- (in $n_0(x)$)
and multiple- (in $n_1(x)$) reflections from each turning point \cite{BM72, LY72, LL75, RB08}. 

\noindent{\em Evanescent:} For $x$ far outside the classically allowed region for the density,
$-z_F(x) >> 1$, and
\ben
\n\sc(x)
\to
\left[\frac{p_F(x)}{3 S_F(x)}- \frac{\omega_F}{p_F(x) \sin\alpha_F(x)} \right]
\frac{e^{-2|S_F(x)|/\hbar}}{4\pi},
\label{evlim}
\een
generalizing the approximation of  Ref. \cite{KSb65}. 
Similarly,
\ben
t\sc(x)
\to
\left[\frac{p_F^3(x)}{3 S_F(x)}- \frac{3\omega_F p_F(x)}{ \text{sin}\alpha_F(x)} \right]
\frac{e^{-2|S_F(x)|/\hbar}}{24\pi},
\label{evtlim}
\een

\noindent{\em Turning point:} At a Fermi energy
turning point $x_0$, where $v'(x_0)\neq 0$, the leading term in the density
is known:
\ben
\n\sc(x_0)
= c_0 \hbar^{-2/3}|dv/dx|^{1/3}, 
\een
where $c_0=  (2/9)^{1/3}/ \Gamma^2(1/3)$
\cite{KSb65}.  In addition,
\ben t\sc(x_0) = - d_0 |dv/dx| \een
where $d_0 = 1/[9 \Gamma(2/3)\Gamma(1/3)]$.

\begin{figure}[t]
\includegraphics[scale=0.5]{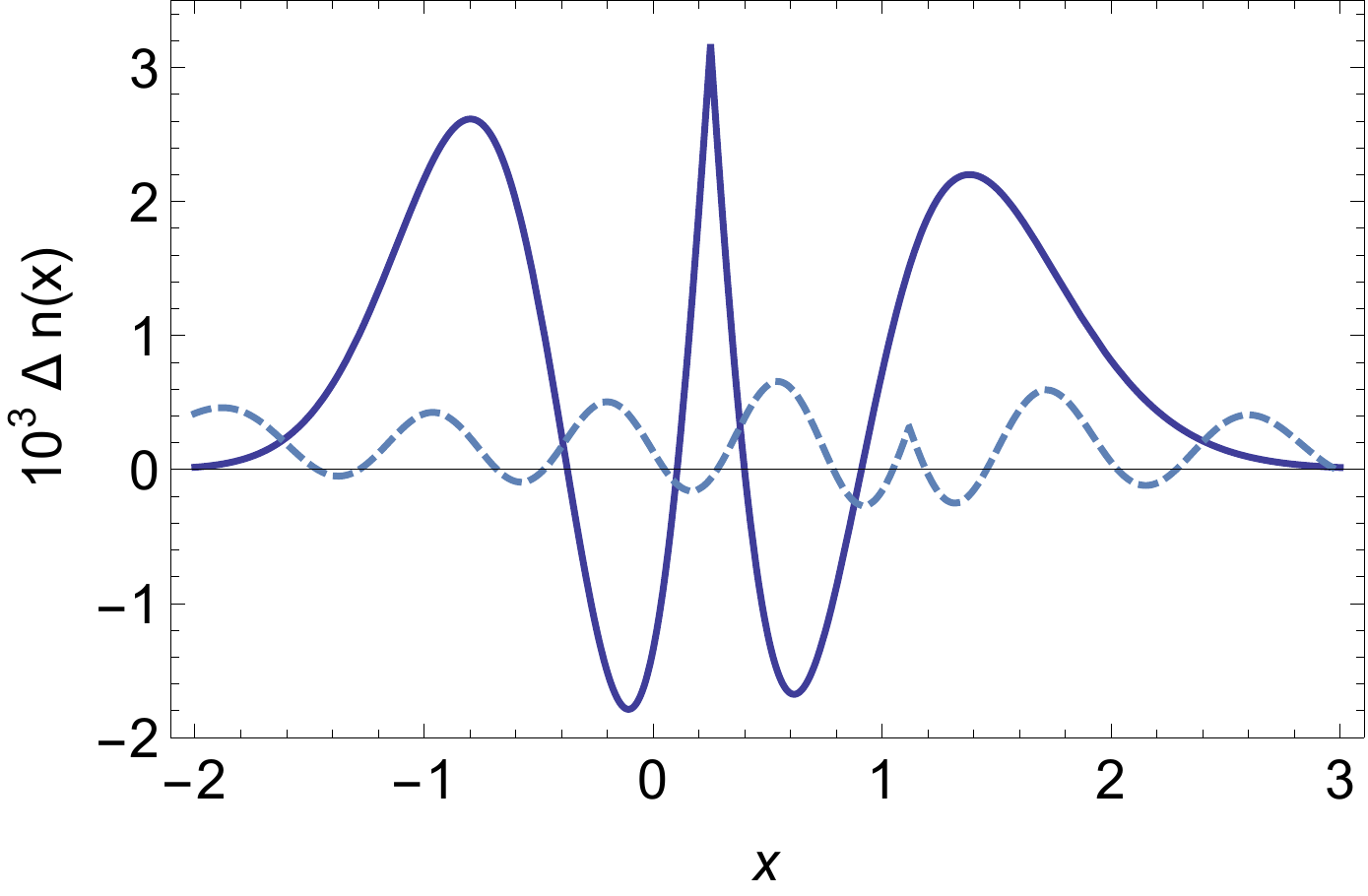}
\caption{Error in 
semiclassical density for
$N=2$ (solid), and $N=8$ (dashed)
in the Morse potential of Fig. 1.}
\label{Morse20DD}
\end{figure}

The present development unifies all earlier partial results\cite{KSb65,LL75,RB08,CLEB10}.
In Fig. \ref{Morse20D}, we showed how accurate the semiclassical
density is in a Morse potential
that supports 21 levels.  In Fig. \ref{Morse20DD}, we
plot the density error for 2 and 8 particles.
The cusp in the center is at the mid-phase point $x_m^{N-1/2}$ where the left- 
meets the right-turning point solution.  

\begin{figure}[hb]
\includegraphics[scale=0.5]{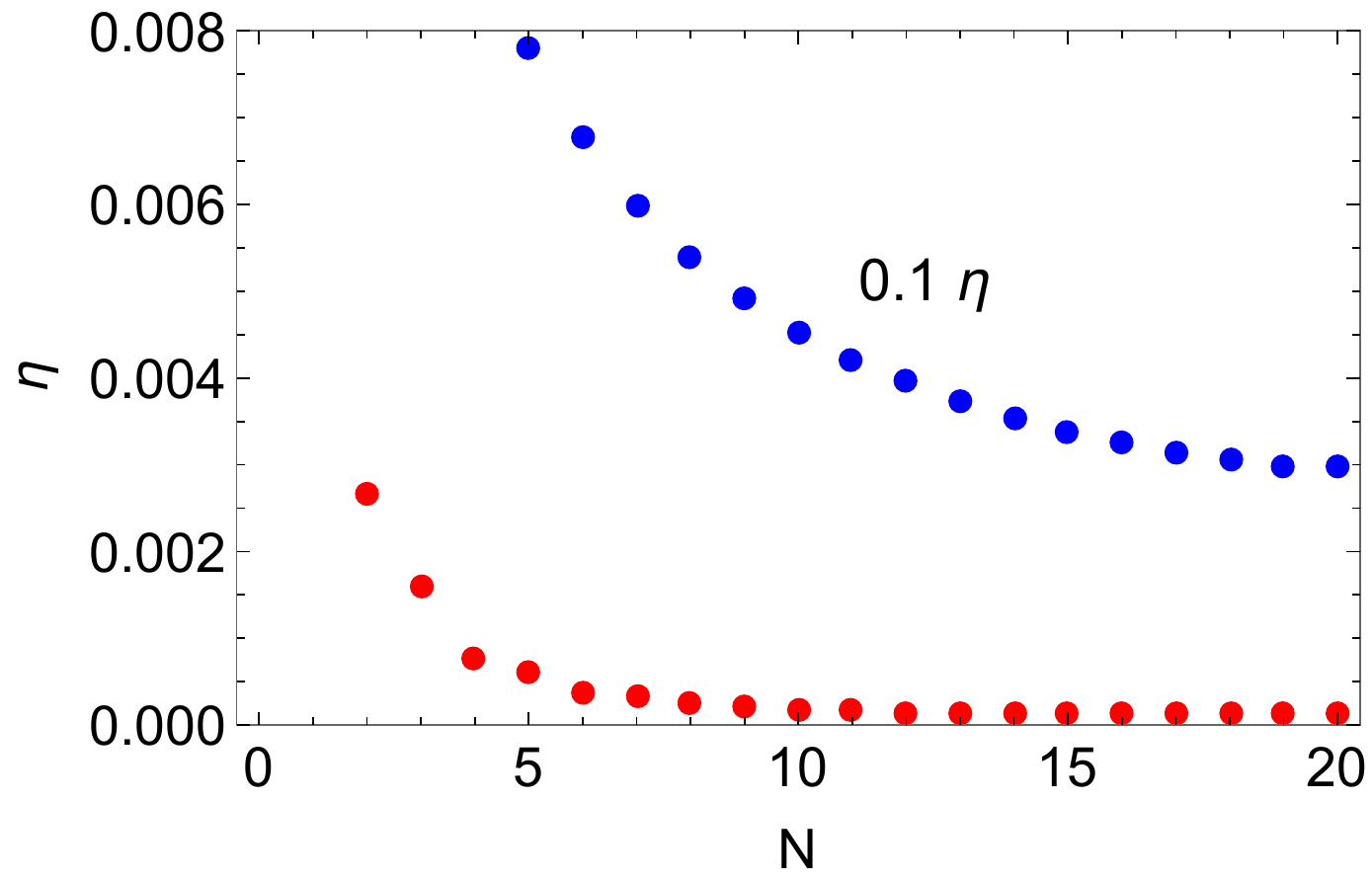}
\caption{Integrated measure of error (Eq. \ref{etadef}) in TF density multiplied by 0.1 (top) and semiclassical
uniform approximation (bottom) for the Morse potential of Fig. 1.} 
\label{Morse20Derror}
\end{figure}

To quantify, we define
a measure of density difference as
\ben
\eta=\frac{1}{N}\int_{-\infty}^{\infty} dx\, |n^{sc}(x) - n(x)|,
\label{etadef}
\een
which only vanishes
when two densities are identical pointwise, and remains comparable in magnitude
to the pointwise difference.
In Fig. \ref{Morse20Derror}, we plot this error measure for the uniform approximation for the
number density in Eq. \ref{nsc} and for the
TF density  (Eq.\ref{ntf})
as a function of $N$.
As $N$ grows,
$\eta$ shrinks until levels close to the
unstable point of the well are included.

\begin{figure}[b]
\includegraphics[scale=0.5]{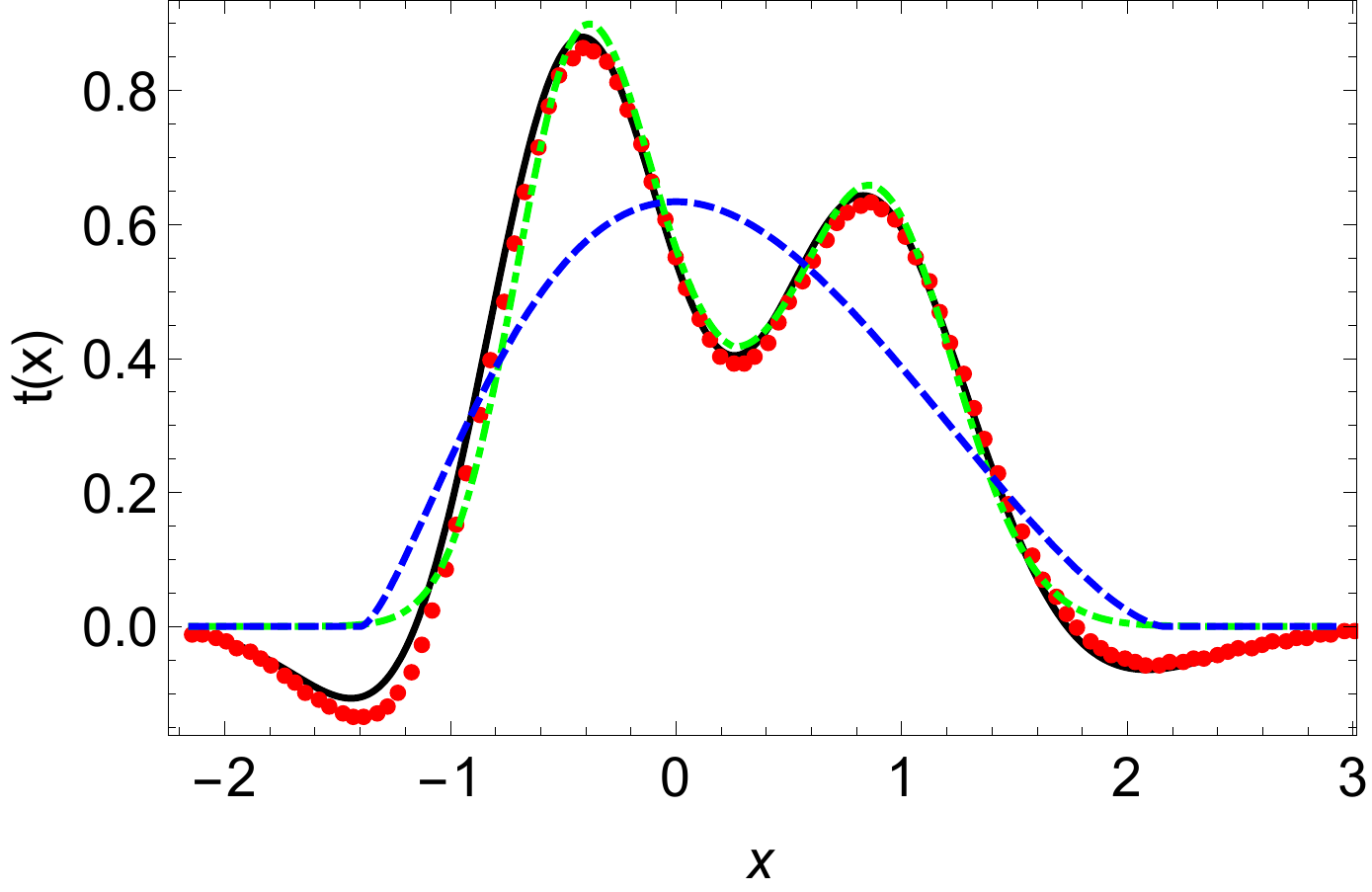}
\caption{Thomas-Fermi (dashed), uniform semiclassical (dotted) and
exact (solid) kinetic energy density for 
2 particles in the Morse potential of Fig. 1.  The value
of $\pi^2 [\n\sc(x)]^3/6$ is also shown (dot-dashed).}
\label{Morse20T}
\end{figure}

\begin{figure}[ht]
\includegraphics[scale=0.5]{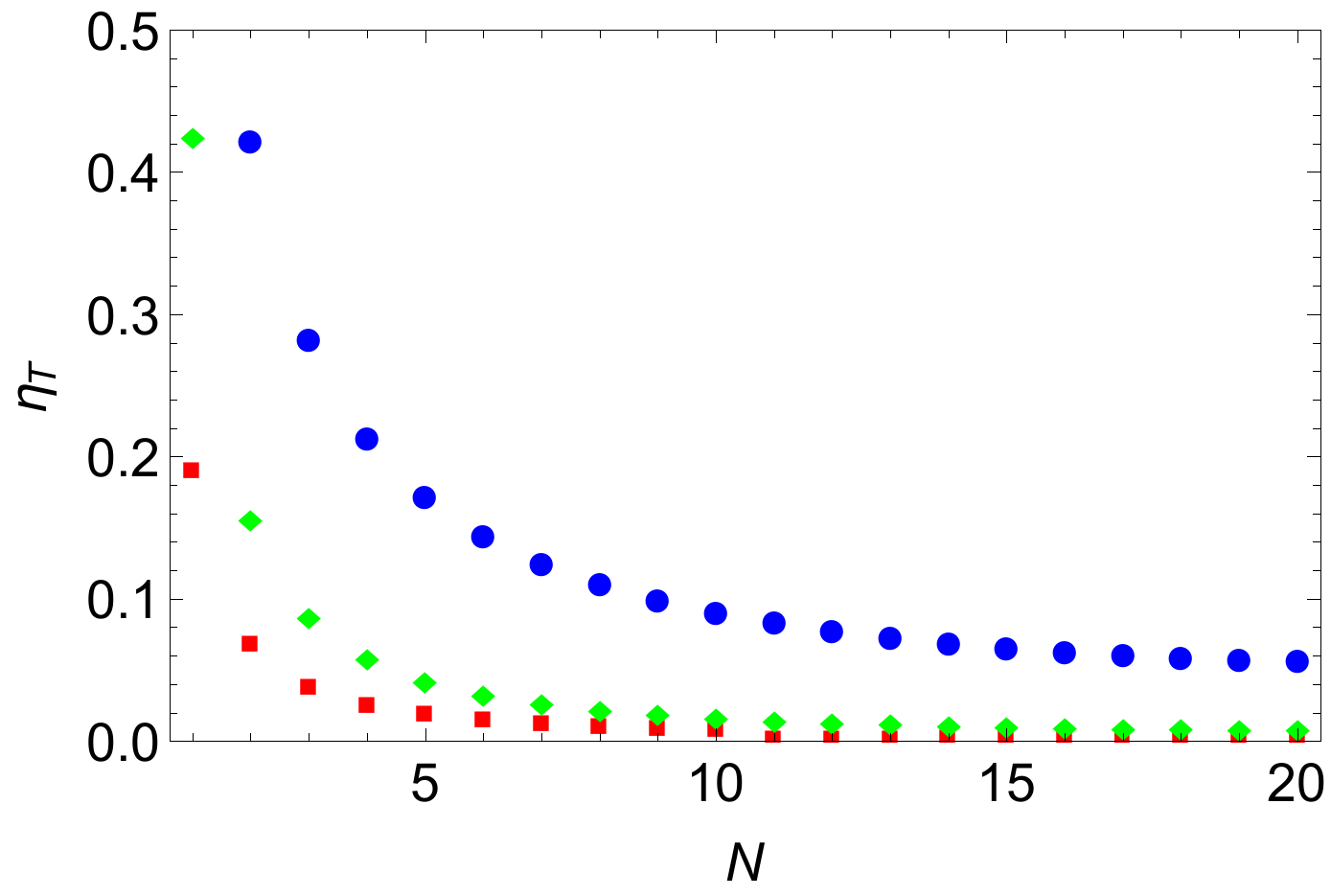}
\caption{Error (see text) in kinetic energy densities in
the Morse potential of Fig. 1 with
the semiclassical uniform approximation (squares), Thomas-Fermi theory (dots)
and $t^{\text{loc}}[n^{\text{sc}}]$ (rhombs).}
\label{Morse20TError}

\end{figure}

In Fig. \ref{Morse20T}, we plot $t(x)$.
The TF result
clearly misses the oscillations and everything beyond the turning
points.  The exact $t(x)$ becomes negative near the turning points and
this effect is well captured by the uniform semiclassical approximation.
Brack et al. \cite{RBK10} noted that
$t^{\text{loc}}[n] = \pi^2 n^3/6$ evaluated on the exact density can yield
an accurate approximation, but only in the classically-allowed region.
The improvement of the uniform approximation with increasing $N$ is
reflected in Fig. \ref{Morse20TError}, in which $\eta_T$ is defined
analogously to Eq. (27) except with the exact $T$ in the denominator.
We find qualitatively similar results for several other systems including
those with uncountable  (Rosen-Morse\cite{RM32} 
potential) and countable spectra (simple harmonic oscillator and quartic
oscillator). 
Longer accounts of the derivation, performance, and relation to 
DFT are in preparation.

Eq. (\ref{nsc}) cannot be applied to three-dimensions, Coulomb potentials,
multi-center problems or interacting particles, whereas
TF theory can be applied to almost any fermionic problem.
But Eq. (\ref{nsc}) strongly suggests corrections to TF exist (even if they
can only be evaluated numerically), are extremely accurate,
and must reduce to Eq. (\ref{nsc}) where
applicable.  Without Eq. (\ref{nsc}), we would have no reason to search for
them.  Now we have.

We acknowledge NSF Grant NO. CHE-1112442.
We thank Michael Berry for useful discussions.



\bibliography{MasterOld}

\end{document}